# Implementation of on-chip multi-channel focusing wavelength demultiplexing with regularized digital metamaterials

*Jie Huang,[†] Junbo Yang,[\*,†,‡] Dingbo Chen,[†] Wei bai,[§] Jingmin Han,[ξ] Zhaojian Zhang,[†] Jingjing Zhang,[†] Xin He,[†] Yunxin Han,[†] and Linmei Liang[†]*

[†]Center of Material Science, National University of Defense Technology, Changsha, Hunan 410073, China

[‡]China State Key Laboratory on Advanced Optical Communication Systems and Networks, Peking University, Beijing 100871, China

[§]The Key Laboratory of Adaptive Optics, Chinese Academy of Sciences, Chengdu, Sichuan 610209, China

[ξ]College of Electronic and Information Engineering, Southwest University, Chongqing 400715, China

*Corresponding Author: yangjunbo@nudt.edu.cn

**ABSTRACT:** Adiabatic waveguide taper and on-chip wavelength demultiplexer are the key components of photonic integrated circuits. However, these two kinds of devices which designed by traditional semi-analytic methods or brute-force search methods usually have large size. Here, based on regularized digital metamaterials, we have designed, fabricated and characterized a two-channel focused wavelength demultiplexer with a footprint of 2.4 × 10 μm$^2$. The designed demultiplexer can directly connect to a grating coupler under the absence of an adiabatic waveguide taper. The objective first method and modified steepest descent method are used to design the demultiplexer which splits 1520 nm and 1580 nm light from a 10-μm-wide input waveguide into two 0.48-μm-wide output waveguides. Experimental results show that the insertion loss of the upper (lower) channel of the demultiplexer is -1.77 dB (-2.10 dB) and the crosstalk is -25.17 dB (-12.14 dB). Besides, the simulation results indicate that the fabrication tolerance of our devices can reach ±20 nm in etching depth and ±10 nm in plane size changing. Benefit From the extensibility of our design method, we can design other types of ultra-compact 'focused' devices, like mode splitters, mode converters and power splitters, and we can also design devices with more complicated functionalities, for example, we have designed a three-channel focused wavelength demultiplexer.

**INTRODUCTION**

For photonic integrated circuits (PIC), high density means low power consumption and high performance. However, conventional components of integrated photonic devices which designed by searching for a relatively small parameter space generally have bigger footprints.[1-3] Driven by the demand of highly integrated photonic circuits, various computational optimization methods have been proposed to decrease the footprints of single devices, including objective first method,[4,5] direct-binary-search (DBS) method,[6,7] topology optimization,[8] adjoint shape optimization,[9-11] genetic algorithm (GA)[12,13] and deep learning.[15,16] Through applying these design methods, numerous of digital-metamaterials-based devices have been proposed,[17] ranging from quantum applications[18] to metalens.[19,20] Nevertheless, the digital metamaterials enabled devices mostly have sole function.[4-16,21-26] To further increase the density of PIC, multi-functional devices like wavelength-division multiplexing (WDM) grating couplers have also been demonstrated.[27] The proposed WDM grating can separate vertically incident light into two separate waveguides with high splitting ratio, but due to the high selectivity of the grating to working wavelengths, it will be difficult to design other complicated WDM gratings, for example, the multi-channel WDM grating and broadband WDM grating.

In order to match the mode field of standard single-mode fiber, the longitudinal size of ordinary grating couplers are often on the scale of 10 μm,[28] more importantly, for the sake of realizing low loss interconnection between single-mode waveguide and 10-μm-wide waveguide, an adiabatic waveguide taper with a length of hundreds of micrometers will be needed.[29] Based on the concept of digital metamaterials, Liu and colleagues have designed a waveguide taper for interconnection between the 0.5-μm-wide single-mode waveguide and the 10-μm-wide multi-mode waveguide.[24] In contrast to ordinary adiabatic waveguide taper, the designed waveguide taper has a relatively small length (5 μm). Unfortunately, limited by the low optimization efficiency of DBS method, it can be time-consuming (about 100 hours) to optimize such a large design area (2952 square pixels). Besides, due to the absence of gradient information, the convergence of multi-objective problem will be relatively slow and the final optimization result is often unacceptable. In this paper, by using objective first method[4,30,31] and modified steepest descent method, we have experimentally demonstrated a two-channel focused wavelength demultiplexer with a footprint of 2.4 × 10 μm$^2$. The word 'focus' means the demultiplexer is directly connected to a 10-μm-wide grating coupler without the assistance of adiabatic waveguide taper. After the broadband input light passed through the demultiplexer, the light at different wavelengths will be coupled out through different 0.48-μm-wide waveguides. Different from our previous work,[31] here, to ensure the fabricability of the designed demultiplexer, we applied the minimum feature size constraint and binarization in the steepest descent method.[11,32] Furthermore, to verify the ability of our design method in designing devices with more complex functions, we have designed a three-channel focused wavelength demultiplexer.

**RESULTS AND DISCUSSION**

**Optimization Model Construction.** Based on the concept that the performance of a linear optical device can be specified by defining the mode conversion efficiency between a set of input and output modes,[33] Jesse and colleagues proposed a universally applicable mode to construct the optimization problem.[4] As schematically shown in Figure 1, for the two-channel focused wavelength demultiplexer, there are two modes input from port 1, the two input modes are fundamental transverse electric (TE) mode at the wavelengths of 1520 nm (mode 1) and 1580 nm (mode 2), respectively. The electric fields $\mathbf{E}_i$ generated by the input mode $i\,(i=1,2)$ should satisfy Maxwell's equations in the frequency domain,[5]

$$\nabla \times \mu_0^{-1} \nabla \times \mathbf{E}_i - \omega_i^2 \varepsilon \mathbf{E}_i = -i\omega_i \mathbf{J}_i \qquad (1)$$

where $\mu_0$ is the permeability of vacuum, $\omega_i$ is the angular frequency of mode $i$, $\varepsilon$ is the space-dependent permittivity and $\mathbf{J}_i$ is the equivalent excitation current density distribution of the input mode $i$.

For each input mode $i$ of the demultiplexer, we can specify $N_i$ output modes, the mode conversion efficiency of input mode $i$ to target output mode $j\,(j=1,...,N_i)$ can be computed by using overlap integrals.[33] Then we can define the device performance by constraining the amplitudes of the mode conversion efficiencies to be between $\alpha_{ij}$ and $\beta_{ij}$, which can be expressed as[5]

$$\alpha_{ij} \le \left| \iint_{S_{ij}} \mathbf{c}_{ij}^\dagger \cdot \mathbf{E}_i dS \right| \le \beta_{ij} \qquad (2)$$

where $c_{ij}$ represents the electric fields of output mode $ij$, $S_{ij}$ represent the surfaces of the output modes, $\alpha_{ij}$ and $\beta_{ij}$ represent the lower and upper limits of the amplitude of mode conversion efficiencies, respectively. The optimization of the demultiplexer now equivalent to finding the space-dependent permittivity $\varepsilon$ and electric fields $\mathbf{E}_i$ that simultaneously satisfy (1) and (2).

As shown in Figure 1, the demultiplexer was designed on a fully etched 220-nm-thick Si $(n_{Si}=3.477)$ layer with 2-μm-thick SiO$_2$ $(n_{SiO_2}=1.445)$ substrate and air $(n_{air}=1)$ cladding. The left side of the demultiplexer is directly connected to a photonic crystal grating through a 10-μm-wide waveguide, which means the fundamental TE mode of the 10-μm-wide waveguide was used as the input mode for the optimization. After passed through the demultiplexer, the input light will be coupled out through two 0.48-μm-wide waveguides, which means the output modes are the fundamental TE mode of two 0.48-μm-wide output waveguides. For port 2, we specified that the output power of fundamental TE mode at 1520 nm should be in the range of [0.9 1] and the output power of fundamental TE mode at 1580 nm should be in the range of [0 0.01]; the converse was specified for port 3.

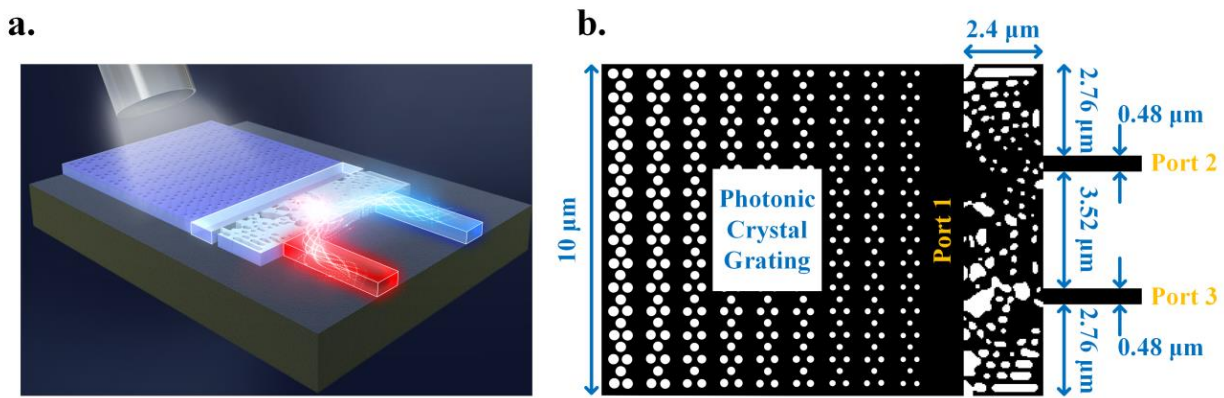

Figure 1. Schematic of the designed two-channel focused wavelength demultiplexer. Three-dimensional diagram (a) and top view (b).

**Binarization and Regularization.** We divide the optimization process into two separate parts, the continuous optimization and discrete optimization which use the objective first method and modified steepest descent method, respectively. In continuous optimization, the design targets in equation (2) are preferentially satisfied, then the 'Alternating Directions Method of Multipliers'

(ADMM) optimization algorithm are employed to minimize the violation of physics in equation (1).[4,5] As shown in Figure 2 (b), by applying the ADMM algorithm, the objective first method can quickly produce a continuously varying permittivity distribution which has distinct gradient.

For a realistic nanophotonic device, the permittivities in the design area have to be a finite number of discrete values, for example, the permittivities of silicon and air. Methods like binary level-set representation has been proposed to implement discrete optimization.[4,5] The proposed method works well if the continuous optimization has already yielded a mostly-discrete device. However, if the yielded permittivity contains large areas where the permittivities are close to the threshold level, the transition from continuous optimization to binary level-set representation will lead to a serious functional degradation.[34] To mitigate this issue, Logan and colleagues proposed a special penalty function (neighbor biasing).[23] Unfortunately, there is still performance degradation when transit from continuous optimization to binary level-set representation. Here, in discrete optimization, we constrain the electric fields $\mathbf{E}_i$ to satisfy Maxwell's equations and define a figure of merit (FOM) function based on the violation of the design targets in equation (2). Then by using the adjoint method, we can compute the gradient of the FOM function with respect to the permittivity parameter $\mathbf{z}$, where $\mathbf{z} \in [0\ 1]^n$ ($n$ is the total number of pixels in the design area) is the permittivity parameter, the permittivity of $m$th pixel is determined by $\varepsilon_m = (\varepsilon_{Si} - \varepsilon_{air})z_m + \varepsilon_{air}$. But for a realistic nanophotonic device, $z_m$ can only be 0 or 1, in order to binarize the device, we use the following projection scheme to recreate the permittivity parameter,[11,32]

$$\overline{z}_m = \frac{\tanh(\beta\eta) + \tanh(\beta[z_m - \eta])}{\tanh(\beta\eta) + \tanh(\beta[1-\eta])} \quad (3)$$

here, $\beta$ controls the strength of the projection, $\eta$ controls the mid-point of the projection. In the discrete optimization here, $\eta$ is fixed at 0.6, $\beta$ is periodically increased from 5 to 20. Now we have $\varepsilon_m = (\varepsilon_{Si} - \varepsilon_{air})\overline{z}_m + \varepsilon_{air}$ and we can use the gradient $\partial\mathbf{FOM}/\partial\varepsilon \cdot \partial\varepsilon/\partial\overline{\mathbf{z}} \cdot \partial\overline{\mathbf{z}}/\partial\mathbf{z}$ to perform steepest descent optimization. Figure 2 (c) shows the result of discrete optimization, where the black pixels represent silicon and white pixels represent air, it clearly shows that the design area of the device has been fully binarized.

Although the above-mentioned projection scheme can produce a fully binarized device, lots of tiny air holes (smaller than 80nm) appeared in the design area, which means the fabrication of the designed device will be difficult. In fact, the unconstrained binary level-set representation method faces the same problem. To solve this problem, Alexander and colleagues proposed fabrication-constrained level-set method.[36] Here, in order to solve this problem, we employed a low pass filter to regularize the pixels in the design area,

$$\tilde{z}_m = \frac{\sum_{k \in D} W_{mk} z_m}{\sum_{k \in D} W_{mk}} \quad (4)$$

where $D$ represents the design area, and $W$ is the low pass filter, which defined a minimum feature size $R$ as

$$W_{mk} = \begin{cases} R - |r_m - r_k| & \text{if } |r_m - r_k| \leq R \\ 0, & \text{otherwise} \end{cases} \quad (5)$$

where $|r_m - r_k|$ is the distance between pixels $m$ and $k$.

A fabricable device should be regularized and binarized, which means the projection scheme and the low pass filter should be combined, so we rewrite equation (3) to following equation,

$$\bar{z}_m = \frac{\tanh(\beta\eta) + \tanh(\beta[\tilde{z}_m - \eta])}{\tanh(\beta\eta) + \tanh(\beta[1-\eta])} \tag{6}$$

now we have the gradient of the FOM function with respect to the permittivity parameter $\mathbf{z}$,

$$\frac{\partial \mathbf{FOM}}{\partial \mathbf{z}} = \frac{\partial \mathbf{FOM}}{\partial \varepsilon} \cdot \frac{\partial \varepsilon}{\partial \bar{\mathbf{z}}} \cdot \frac{\partial \bar{\mathbf{z}}}{\partial \tilde{\mathbf{z}}} \cdot \frac{\partial \tilde{\mathbf{z}}}{\partial \mathbf{z}} \tag{7}$$

perform steepest descent optimization with this gradient, we can obtain the optimized demultiplexer which is composed of regularized digital metamaterials. In our optimization, $\eta$ is fixed at 0.6, $\beta$ is periodically increased from 5 to about 100, $R$ are fixed at 3 (Figure 2 (d)) or 4 (Figure 2 (e)). As shown in figure 2 (d-e), the combination of projection and low pass filter can binarize the pixels in the design area and remove the small features at the same time. We can also find that the minimum feature size will increase with the increase of $R$. However, as illustrated in Supporting Information, the performance degradation of the devices also increase with the enhancement of the constrains, which means we need to balance the performance and fabricability of the designed devices.[37,38] Here, we chose the minimum feature size $R$ to be 3 (equivalent to 120 nm).

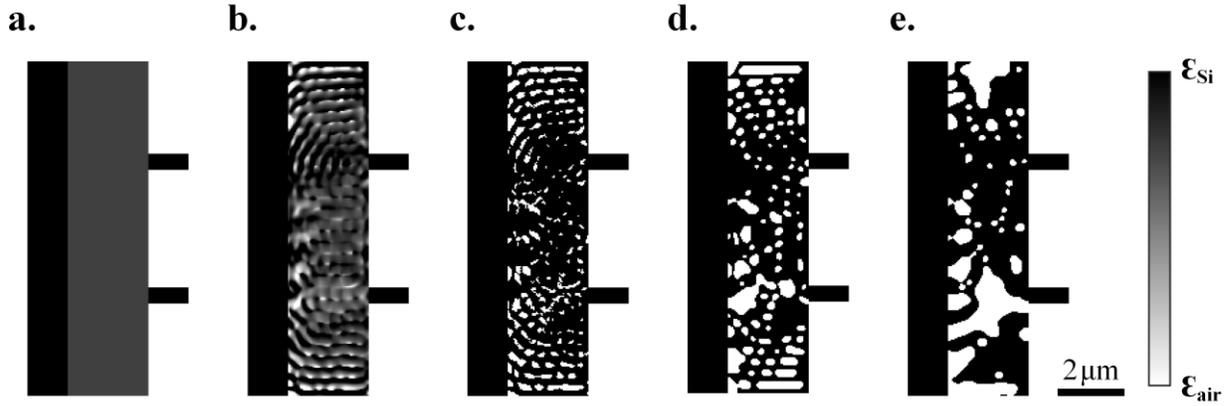

Figure 2. Structures of the designed focused wavelength demultiplexers, where black represents silicon and white represents air. (a). Initial structure, where all the values of the permittivity parameter **z** in the design area are set to 0.75. (b). Continuous optimized structure. (c). Discrete optimized structure which obtained by applying projection alone. (d-e). Discrete optimized structures which obtained by applying projection and low pass filter at the same time. The minimum feature size $R$ are set to be 3 (d) and 4 (e), respectively.

**Experimental Results and Analyses.** Figure 3 (a) shows the scanning electron microscopy (SEM) image of the fabricated focused wavelength demultiplexer with minimum feature size $R=3$. The image show that all of the designed air holes were precisely reproduced by the fabrication process, which means constraining the minimum feature size to 120 nm is acceptable for a realistic device. Figure 3 (b) are the simulated electromagnetic energy density distributions $\left(\frac{1}{2}\varepsilon_0|\mathbf{E}|^2 + \frac{1}{2}\mu_0|\mathbf{H}|^2\right)$, which obtained by using Lumerical FDTD Solutions.[39] It clearly indicated

that the input light at different wavelengths can be focused and coupled out through different output waveguides.

As depicted in Figure 4, the simulated peaking transmission was -1.39 dB at 1507 nm, -1.45 dB at 1565 nm, the corresponding crosstalk was -26.69 dB and -23.50 dB. Figure 4 also shows the transmission of the fabricated device, the peaking transmission was -1.77 dB at 1524 nm, -2.10 dB at 1584 nm, the corresponding crosstalk was -25.17 dB and -12.14 dB. The discrepancies between simulated and measured devices were resulted from the fabrication imperfections. More specifically, due to the influence of the reactive ion etching (RIE) lag effect,[40] under-etching will occur in the regions with relatively small feature size. The under-etching of these small holes can cause device function degradation and red shift of the transmission spectrum. Besides, as shown in Figure 3 (b), in contrast to that at the wavelength of 1507 nm, the density of the electromagnetic energy at the wavelength of 1565 nm is relatively strong in almost the entire design region. Therefore, the under-etching of these small holes affects the performance of lower channel more than it affects the performance of upper channel, which means for the fabricated devices, the performance degradation of the lower channel will worse than that of the upper channel.

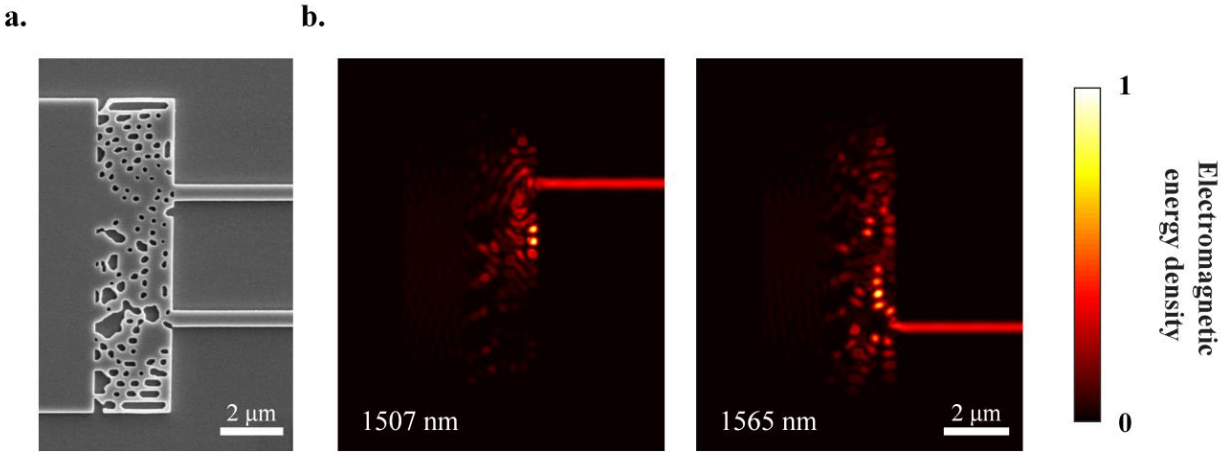

Figure 3. Two-channel focused wavelength demultiplexer with minimum feature size $R = 3$. (a) SEM image of the fabricated device. (b) Simulated electromagnetic energy density in a horizontal slice through the middle of the device, the wavelengths are 1507 nm and 1565 nm, respectively.

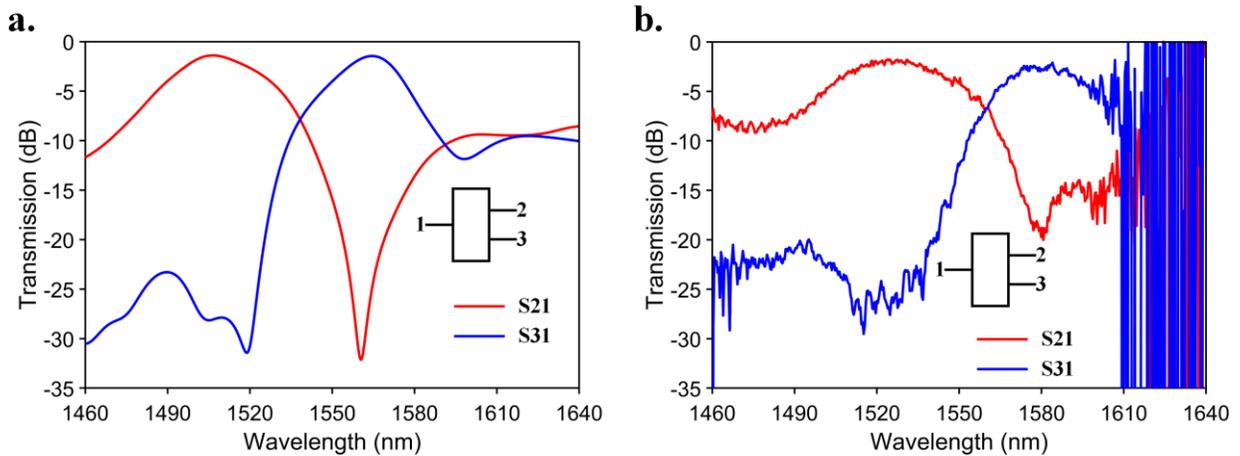

Figure 4. Simulated and measured S-parameters for the focused wavelength demultiplexer, where $S_{ij}$ is the transmission from port $j$ to port $i$. (a) Simulated S-parameters by using Lumerical FDTD Solutions. (b) Measured S-parameters of the fabricated demultiplexer.

**Tolerance to Fabrication Errors.** Figure 5 shows the robust analysis to etching depth and plane size changing of the air holes in the design area. Here, to simplify the problem, we assume that the depth and plane size of all air holes vary uniformly. As shown in Figure 5 (a), under-etching will cause function degradation and the degradation of the lower channel is bigger than that of the upper channel, this conclusion is consistent with our previous analysis (Figure 4). Besides, Figure 5 (a) also indicated that the under-etching of the devices can cause red shift of the transmission spectrum. As described in our previous work,[41] this is because a shallower etch depth will lead to a bigger equivalent refractive index in the design area. Conversely, due to the TE mode is mostly confined to the silicon layer, the over-etching of the air holes has small influence on the device performance.[6] Figure 5 (b) shows that change the plane size of the air holes can result in function degradation too, and increase (decrease) the plane size of the air holes can cause the blue (red) shift of the transmission spectrum. As described in Figure 5, even for the worst case (plane size changing of +10 nm), the peaking transmissions of the upper (lower) channel was -2.63 dB (-2.41 dB) and the corresponding crosstalk was -18.56 dB (-13.85 dB), which means if the peaking transmissions of the focused demultiplexer are allowed to drop 1.3 dB from the peak, then the etch depth (plane size) of the etched holes can vary by as much as ±20 nm (±10 nm).

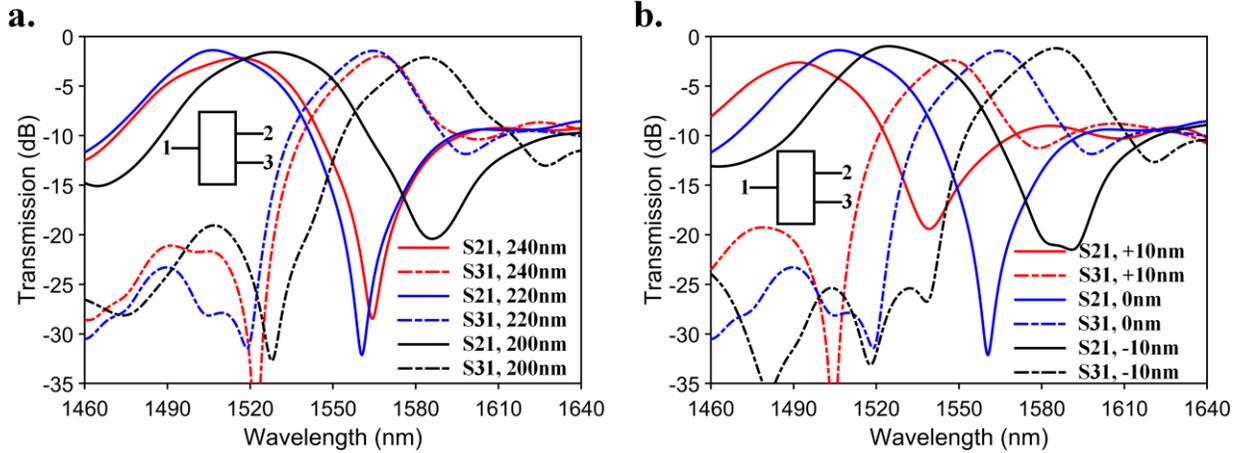

Figure 5. Robust analysis to fabrication errors. (a) Simulated S-parameters with etching depth to be 240 nm (red), 220 nm (blue) and 200 nm (black). (b) Simulated S-parameters with plane size changing of +10 nm (red), 0 nm (blue) and -10 nm (black). The solid (dashed) line represent the S-parameters of the upper (lower) channel.

**Three-Channel Focused Wavelength Demultiplexer.** To illustrate the ability of our method to design devices with more complex functions, we have numerically designed a three-channel focused wavelength demultiplexer. The width of the demultiplexer is consistent with the width of the two-channel focused wavelength demultiplexer (10 μm). In order to make sure the design process can coverage to an acceptable result, we increased the length of the demultiplexer from 2.4 μm to 4.8 μm. As described in Figure 6 (b), the simulated peaking transmission was -1.36 dB at 1477 nm, -0.95 dB at 1537 nm, -1.11 dB at 1597 nm, the corresponding crosstalk was under -23.68 dB, -25.89 dB and -17.41 dB. The simulated electromagnetic energy density in Figure 6 (c) indicated that the input broadband light from 10-μm-wide waveguide can be focused and coupled out through different 0.48-μm-wide output waveguides.

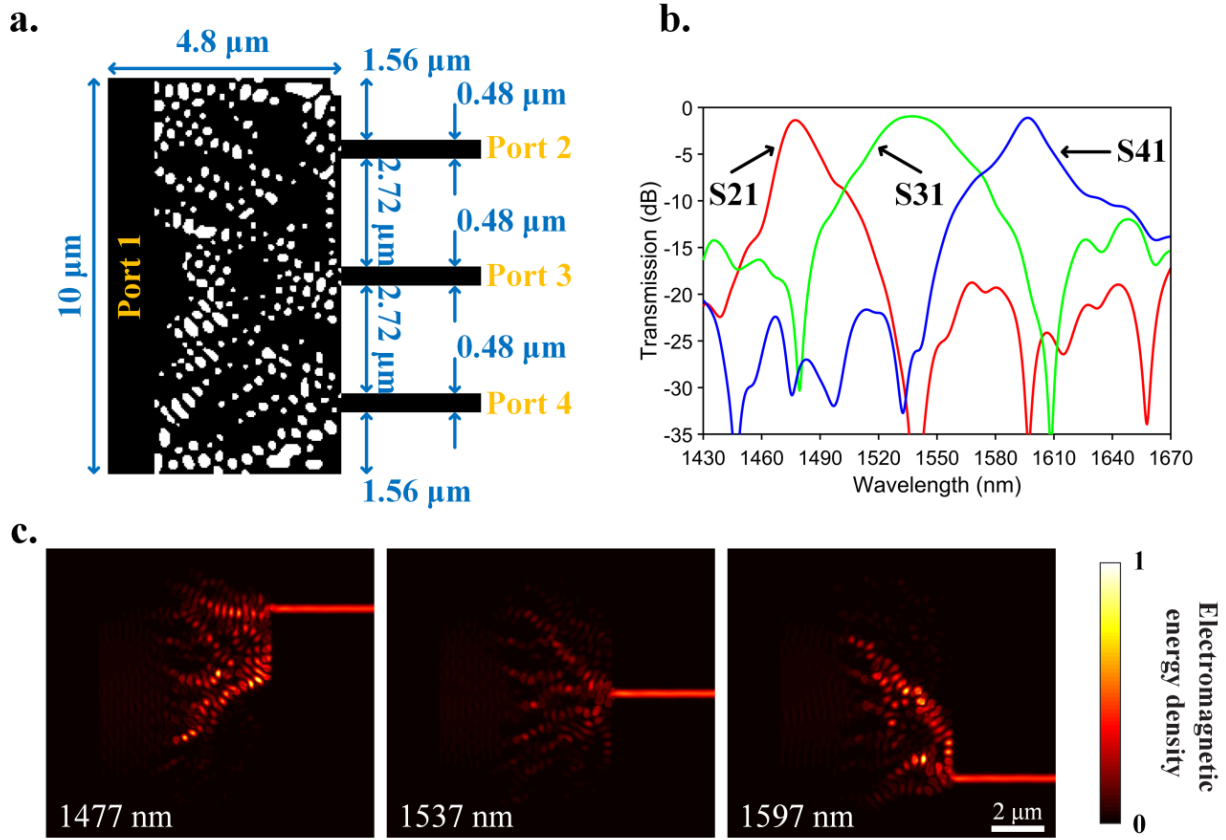

Figure 6. Three-channel focused wavelength demultiplexer with minimum feature size $R = 3$. (a) Structures of the designed demultiplexer, where black represents silicon and white represents air. The length (width) of the demultiplexer is 4.8 μm (10 μm). (b) Simulated S-parameters for the demultiplexer, where $S_{ij}$ is the transmission from port $j$ to port $i$. (c) Simulated electromagnetic energy density in a horizontal slice through the middle of the demultiplexer, the wavelengths are 1477 nm, 1537 nm and 1597 nm, respectively.

**CONCLUSIONS**

In conclusion, based on the concept of regularized digital metamaterials, we have designed, fabricated and characterized a two-channel focused wavelength demultiplexer with a footprint of

2.4 × 10 μm². The designed demultiplexer can realize mode conversion and wavelength splitting in the absence of long adiabatic waveguide taper. By using objective first method, we can obtain a good initial permittivity distribution of the design area which has distinct gradient, then by using modified steepest descent method, we can regularize and binarize the permittivity distribution, which will produce a fabricable device. The peaking transmission of the fabricated device was -1.77 dB at 1524 nm, -2.10 at 1584 nm, and the corresponding crosstalk was -25.17 dB and -12.14 dB. Furthermore, the simulation results indicated that the designed device is very robust to fabrication errors. The experimentally demonstrated design method can also be applied to design other types of 'focused' devices and design devices with more complicated functionality, for example, we have numerically designed a three-channel focused wavelength demultiplexer.

## METHODS

**Fabrication.** The two-channel focused wavelength demultiplexers were fabricated on a silicon-on-insulator (SOI) wafer with 220-nm-thick top silicon layer and 2-μm-thick buried oxide layer. A Vistec EBPG 5000 Plus electron-beam lithography (EBL) system was used to define the optimized nanohole patterns. These patterns were then transferred to the top silicon layer by using an Oxford Plasmalab System100 inductively coupled plasma (ICP) etcher.

**Characterization.** Here, the transmission of the fabricated demultiplexers are measured by using a spectrum analyzer (Yokogawa AQ6370D) with a supercontinuum light source (NKT Photonics SuperK Compact). A series of fully etched photonic crystal grating couplers were fabricated at each input and output port for chip-fiber coupling.

ACKNOWLEDGMENTS

This work was supported by National Natural Science Foundation of China (60907003, 61805278, 61875168), China Postdoctoral Science Foundation (2018M633704, 2017M612885), Foundation of NUDT (JC13-02-13, ZK17-03-01) and Hunan Provincial Natural Science Foundation of China (13JJ3001). We thank Dr. Cheng Zeng for his help in fabrication, Dr. Wei Peng for his help in algorithm development and Dr. Yuedi Ding for her guidance in characterization, we also thank Dr. Yuxi Wang, Dr. Weijie Chang, Dr. Tingan Li and Dr. Long Zhang for helpful discussions. We thank the National Supercomputing Center in Guangzhou (NSCC-GZ) for providing computing resource. We thank all the engineers in the Center of Micro-Fabrication and Characterization (CMFC) of WNLO for the supporting in device fabrication.

SUPPORTING INFORMATION

1. **Constraints on minimum feature size.**

As illustrated in Figure S1, when the value of the minimum feature size $R$ was set to 3 (4), the simulated peaking transmission was -1.39 dB (- 2.25 dB) at upper channel, -1.45 dB (-2.15 dB) at lower channel, the corresponding crosstalk was -26.69 dB (-29.17 dB) and -23.50 dB (-17.65 dB). The simulation results indicate that the performance of the device will degrade with the enhancement of the minimum feature size constrain.

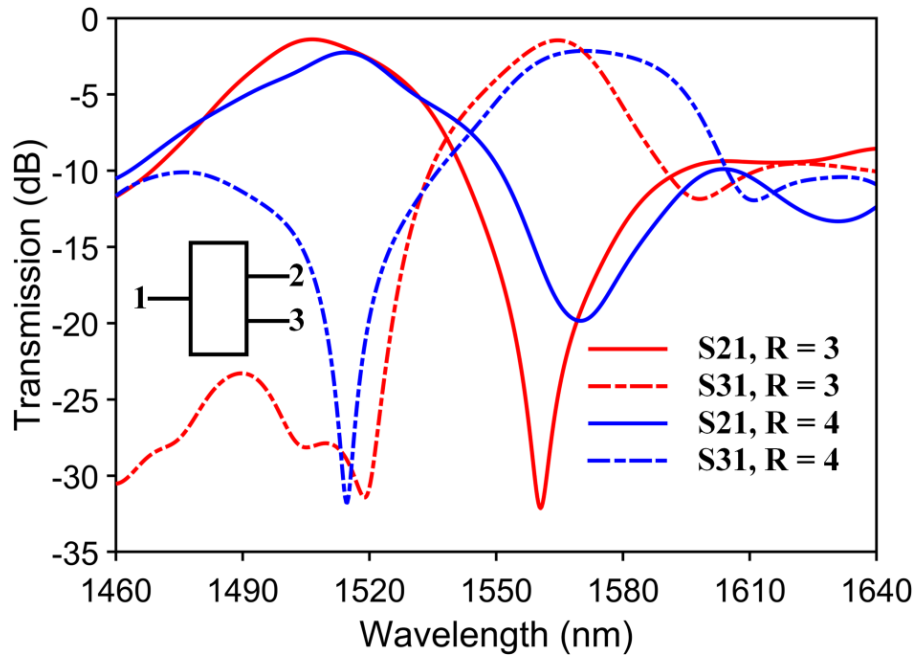

Figure S1. Simulated S-parameters with minimum feature size $R$ to be 3 (red) and 4 (blue). The solid (dashed) line represent the S-parameters of upper (lower) channel.